\documentclass[11pt,a4paper]{article}
\pdfoutput=1

\usepackage{amsmath,amssymb,tikz}
\usepackage{graphicx}
 \allowdisplaybreaks
\def\be{\begin{equation}}
\def\ee{\end{equation}}
\def\ba#1\ea{\begin{align}#1\end{align}}
\def\bg#1\eg{\begin{gather}#1\end{gather}}
\def\bm#1\em{\begin{multline}#1\end{multline}}
\def\bmd#1\emd{\begin{multlined}#1\end{multlined}}

\def\a{\alpha}

\def\c{\chi}

\def\e{\epsilon}

\def\la{\label}

\def\({\left(}
\def\){\right)}
\def\[{\left[}
\def\]{\right]}

\def \be {\begin{equation}}
\def \ee {\end{equation}}
\def \ba {\begin{array}}
\def \ea {\end{array}}
\def \bea{\begin{eqnarray}}
\def \eea{\end{eqnarray}}

\def \a {\alpha}

\def \c {\gamma}

\def \e {\epsilon}

\def \la {\leftarrow}
\def \ra {\rightarrow}

\def\bea{\begin{eqnarray}}
\def\eea{\end{eqnarray}}
\newcommand{\eq}[1]{(\ref{#1})}
\newcommand{\bit}{\begin{itemize}}  \newcommand{\eit}{\end{itemize}}
\newcommand{\ben}{\begin{enumerate}}  \newcommand{\een}{\end{enumerate}}

\def\la{\langle}
\def\ra{\rangle}

\long\def\symbolfootnote[#1]#2{\begingroup%
\def\thefootnote{\fnsymbol{footnote}}\footnote[#1]{#2}\endgroup}

\newcommand{\sysu}{{\it School of Physics and Astronomy, Sun Yat-Sen University, 2 Daxue Road, Zhuhai 519082, China}}

\begin{document}
\thispagestyle{empty}
\begin{center}

~\vspace{20pt}

{\Large\bf Holographic BCFT with Dirichlet Boundary Condition}

\vspace{25pt}

Rong-Xin Miao ${}$\symbolfootnote[1]{Email:~\sf
  miaorx@mail.sysu.edu.cn}

\vspace{10pt}${}$\sysu

\vspace{2cm}

\begin{abstract}

Neumann boundary condition plays an important role in the initial proposal of holographic dual of boundary conformal field theory, which has yield many interesting results and passed several non-trivial tests. In this paper, we show that Dirichlet boundary condition works as well as Neumann boundary condition. For instance,  it includes AdS solution and obeys the g-theorem. Furthermore, it can produce the correct expression of one point function, the boundary Weyl anomaly and the universal relations between them. We also study the relative boundary condition for gauge fields, which is the counterpart of Dirichlet boundary condition for gravitational fields. Interestingly, the four-dimensional Reissner-Nordstr$\ddot{\text{o}}$m black hole with magnetic charge is an exact solution to relative boundary condition under some conditions. This holographic model predicts that a constant magnetic field in the bulk can induce a constant current on the boundary in three dimensions. We suggest to measure this interesting boundary current in materials such as the graphene.

\end{abstract}

\end{center}

\newpage
\setcounter{footnote}{0}
\setcounter{page}{1}

\tableofcontents

\section{Introduction}

\begin{figure}[t]
\centering
\includegraphics[width=5cm]{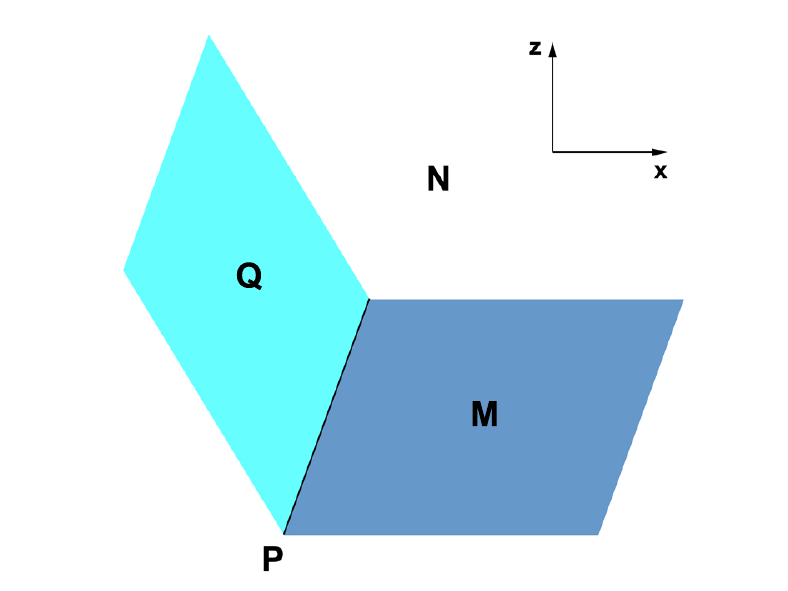}
\caption{Geometry of holographic BCFT}
\label{MNPQ}
\end{figure}

BCFT is a conformal field theory defined on a manifold $M$ with a
boundary $P$ where suitable boundary conditions (BC) are imposed \cite{Cardy:2004hm, McAvity:1993ue}. It has important
applications in quantum field theory, string theory and condensed matter physics. For interesting developments of BCFT and related topics please see
\cite{
  Fursaev:2015wpa,Herzog:2015ioa,Miao:2017aba,Herzog:2017kkj,
Jensen:2017eof, 
Kurkov:2017cdz,Kurkov:2018pjw,Vassilevich:2018aqu,Rodriguez-Gomez:2017kxf,Seminara:2017hhh, Erdmenger:2015spo,Erdmenger:2014xya,Flory:2017ftd,Chu:2018ksb,Chu:2018ntx,Miao:2018dvm,Chu:2018fpx}. In the spirit of AdS/CFT
\cite{Maldacena:1997re}, Takayanagi \cite{Takayanagi:2011zk}
proposes to extend the $d$ dimensional manifold $M$ to a $d+1$
dimensional asymptotically AdS space $N$ so that $\partial N= M\cup
Q$, where $Q$ is a $d$ dimensional manifold which satisfies $\partial
Q=\partial M=P$. See figure \ref{MNPQ} for the geometry. We remark that, unlike $M$, the bulk boundary $Q$ is not the conformal boundary of the manifold $N$. In particular, it is located at finite position instead of infinity. A central issue in the construction of the
AdS/BCFT is the determination of the location of Q in the bulk.  \cite{Takayanagi:2011zk} propose to use the
Neumann boundary condition to fix the position of Q. The holographic BCFT with Neumann BC has produced many elegant results and passed several non-trivial tests \cite{
Nozaki:2012qd,Fujita:2011fp}.  In particular, it obeys the universal relations between Casimir effects and Weyl anomaly \cite{Miao:2017aba}.

In general, there are more than one consistent BCs for a theory. For example, 
one can impose either Neumann (Robin) BC or Dirichlet BC for scalars \cite{McAvity:1990we}
\begin{equation}\label{BCscalar}
\begin{split}
&\text{Neumann (Robin) BC} : (\partial_n + \psi)\phi|_{\partial N}=0,\\
&\text{Dirichlet BC} : \ \ \ \ \ \ \ \ \ \ \ \ \ \phi|_{\partial N}=0.
\end{split}
\end{equation}
Similarly, both absolute BC and relative BC work well for Maxwell fields \cite{Fursaev:2015wpa}
\begin{equation}\label{BCVector}
\begin{split}
&\text{Absolute BC} : \mathcal{F}_{ni}|_{\partial N}=0,\\
&\text{Relative BC} : {}^*\mathcal{F}_{ni}|_{\partial N}=0,
\end{split}
\end{equation}
where $n$ denote the normal direction, $i$ denote the tangent direction and ${}^*\mathcal{F}$ is the Hodge dual of the field strength $\mathcal{F}$.  It is remarkable that absolute BC and relative BC correspond to Neumann BC and Dirichlet BC, respectively \cite{Herzog:2017xha}. Let us explain this in more details below. Consider the variation of the action of Maxwell fields and focus on the boundary terms, we get 
\begin{equation}\label{dIin}
\delta I =-\int_{\partial N} \sqrt{h} \mathcal{F}_n^{\ i}\delta \mathcal{A}_i
\end{equation}
For a well-defined action principle, one can impose either Neumann BC
\begin{equation}\label{NBCV}
\mathcal{F}_{ni}|_{\partial N}=0,
\end{equation}
or  Dirichlet BC such as
\begin{equation}\label{DBCV}
\mathcal{A}_i |_{\partial N}=0 ,
\end{equation}
up to some gauge transformations. 
It is clear that Neumann BC (\ref{NBCV}) corresponds to absolute BC of (\ref{BCVector}). Now let us discuss the Dirichlet BC (\ref{DBCV}). Note that $\mathcal{A}_i$ is defined up to some boundary gauge transformations, i.e., $\mathcal{A}_i \sim \mathcal{A}_i+\partial_i \alpha $.  Instead of using (\ref{DBCV}) directly, it is more convenient to impose its gauge-invariant form 
\begin{equation}\label{DBCV1}
\mathcal{F}_{ij} |_{\partial N}=0,
\end{equation}
which is equivalent to
\begin{equation}\label{DBCV2}
{}^*\mathcal{F}^{ni} |_{\partial N}=\lambda \epsilon^{ni...jk }\mathcal{F}_{jk}|_{\partial N}=0.
\end{equation}
Here $\lambda$ is an unimportant constant,  $n$ denote the normal direction and $(i, ..., j,k )$ denote the tangent directions. Now it is clear that relative BC (\ref{DBCV2}) indeed corresponds to Dirichlet BC (\ref{DBCV}, \ref{DBCV1}).

For gravity, there are also several possible choices of BCs such as Dirichlet BC \cite{Gibbons:1976ue,York:1972sj}, Neumann BC and conformal BC \cite{York:1972sj,Anderson:2010ph,Witten:2018lgb}. As we have mentioned above, the Neumann BC plays an important role  in the construction of holographic BCFT \cite{Takayanagi:2011zk}. As a key characteristic, the BC for holographic BCFT should not only select the solutions to Einstein equations but also determine the location of bulk boundary $Q$. In other words, the position of bulk boundary $Q$ cannot be freely choosen but determined by the BC of gravity. This is indeed the case for Neumann BC \cite{Takayanagi:2011zk}. 

A natural question is that, ``Can Dirichlet BC do the same job for holographic BCFT?''.  Recall that usually one imposes Dirichlet BC on the AdS boundary $M$ in AdS/CFT. Thus it is natural to impose Dirichlet BC on the bulk boundary $Q$ too. The key problem one needs to clarify is that, could Dirichlet BC fix the bulk boundary Q and yield consistent results for holographic BCFT? In this paper, we give a positive answer to this question. We find that AdS is a vacuum solution to Dirichlet BC and the g-theorem is obeyed by this theory.  What is more, Dirichlet BC can also produce the correct forms of one point function, the boundary Weyl anomaly and the universal relations between them. It is remarkable that the central charge for Dirichlet BC is less than or equal to that for Neumann BC. We also study Dirichlet-like BC for gauge fields, the relative BC, and find an exact solution in four dimensions. Interestingly, the holographic theory predicts that there is a constant current on the boundary, when a constant magnetic field is applied in the bulk. And the boundary current gains the maximum value at zero temperature.
It should be mentioned that we have used many methods of  \cite{Miao:2017aba,Chu:2018ntx} in this paper.  \cite{Miao:2017aba,Chu:2018ntx} have developed interesting approach to solve solutions to Neumann BC, and we apply these methods to discuss holographic BCFT with Dirichlet BC in this paper.

The paper is organized as follows. 
In section 2, to warm up, we study the BCs of gauge fields in holographic BCFT. We show that both the absolute BC and the relative BC work well for gauge fields. In section 3, we generalize our discussions to gravitational fields and show that, similar to Neumann BC,  Dirichlet BC can also produce the correct one point function of stress tensor and universal relations between Casimir effects and Weyl anomaly. In section 4, we study the properties of Casimir coefficients and central charges for different BCs. Finally, we conclude with some open questions in section 5.

\section{Boundary Conditions for Maxwell Fields}

In this section we study the BCs of Maxwell fields in holographic BCFT. This can be regarded as a warm-up for the gravitational case. We find that both absolute BC and relative BC can yield the expected asymptotic behaviors of one point function of current near the boundary. Furthermore, both BCs can reproduce the universal relations between the current and Weyl anomaly. 

\subsection{Current and Weyl anomaly}

For the convenience of readers, let us briefly review the Weyl anomaly induced current for BCFT \cite{Chu:2018ksb,Chu:2018ntx}.  In general, the renormalized current of BCFT is singular near the boundary and takes the asymptotic form
\begin{equation}\label{current1}
<J_i>=\frac{\beta_d F_{in}}{x^{d-3}}+O(\frac{1}{x^{d-4}}),\ \ \  x\sim 0,
\end{equation}
where $x$ is the distance to the boundary, $d$ are the dimensions of spacetime,  $F_{ij}$ is a background field strength and $n_i$ is inward-pointing normal vector. It is remarkable that $\beta_d $ are determined by the central charges of Weyl anomaly. For example, we have
\begin{equation}\label{relationa1}
\beta_4=4 b_1, \ \beta_5=2 b_2
\end{equation}
where $b_1, b_2$ are defined by the Weyl anomaly of 4d BCFT and 5d BCFT respectively,
\begin{eqnarray}\label{Weylanomaly1}
&&\mathcal{A}=\int_M \sqrt{g} [b_1 F_{ij} F^{ij}+ \text{curvature terms}], \\
&&\mathcal{A}=\int_{\partial M} \sqrt{h} [b_2 F_{an} F^{an}+b_3 F_{ab}F^{ab} + \text{(extrinsic) curvature terms}]. \label{Weylanomaly2}
\end{eqnarray}
Notice that, since $b_1$ is the bulk central charge, it is independent of the choice of BCs. As a result, the near-boundary current (\ref{current1}) is universal in four dimensions. One the other hand, $b_2$ is the boundary central charge, which implies that the current depends on BCs in five dimensions. Finally, it should be mentioned that there are boundary contributions to the current which can exactly cancel the  apparent ``divergence'' in the bulk current (\ref{current1}) at $x=0$ and define a finite total current \cite{Chu:2018ksb}.

\subsection{Probe limit}

In \cite{Chu:2018ntx}, we have proved that the absolute BC can produce the expected form of near-boundary current (\ref{current1}) and the universal relations (\ref{relationa1}). In this section, we show that this is also the case for relative BC. What is new for relative BC is that there are non-trivial contributions to the current from the bulk boundary $Q$. Such contributions are important in order to recover the universal relations (\ref{relationa1}).

We start with the gravitational action for holographic BCFT
\begin{eqnarray}\label{action}
  I=\int_N \sqrt{G} \Big(
  R-2 \Lambda-\frac{1}{4}\mathcal{F}_{\mu\nu}\mathcal{F}^{\mu\nu}
  \Big)
  +2\int_Q \sqrt{\gamma} (K-T),
\end{eqnarray}
where $\mathcal{F}=d\mathcal{A}$ is the bulk field strength which reduces  to $F=dA$
on the boundary $M$, $K$ is the extrinsic curvature on $Q$ and
$T$ is a constant parameter which can be regarded as the holographic dual
of boundary conditions of BCFT
\cite{Miao:2017gyt,Chu:2017aab}.
Takayanagi \cite{Takayanagi:2011zk} proposes to impose the Neumann BC on $Q$
\begin{eqnarray}
K_{\alpha\beta}-(K-T)\gamma_{\alpha\beta} =0, \label{NBCg}
\end{eqnarray}
which can not only fix the location of boundary $Q$ but also the bulk metrics. In other words, not all the solutions to Einstein equations are allowed in holographic BCFT \cite{Takayanagi:2011zk}. Instead, they have to satisfy the Neumann BC. Ignoring the bulk Maxwell fields, \cite{Takayanagi:2011zk} find AdS 
\begin{equation}\label{AdSmetric}
ds^2=\frac{dz^2+dx^2+\delta_{ab}dy^ady^b}{z^2},
\end{equation}
is a solution to the Neumann BC (\ref{NBCg}), provided that the embedding function of $Q$ is given by
\begin{equation}\label{Q}
\ x=-\sinh \rho\ z,
\end{equation}
where we have re-parametrized $T=(d-2) \tanh \rho$.

Now let us turn to discuss BCs of Maxwell fields. 
In general, there are two consistent BCs on $Q$ for Maxwell fields, i.e., the absolute (electric) BC and relative (magnetic) BC
\begin{eqnarray}
&&\text{Absolute BC} : N^{\mu}\mathcal{F}_{\mu\nu}=0,\label{NBCa}\\
&&\text{Relative BC} :N^{\mu} {}^*\mathcal{F}_{\mu\nu}=0.\label{DBCa}
\end{eqnarray}
Here $N^{\mu}$ is the normal vector on $Q$ and ${}^*\mathcal{F}$ is the Hodge dual of $\mathcal{F}$. The absolute (electric) BC is the Neumann BC, since it fixes the current on the boundary. On the other hand, relative (magnetic) BC is the Dirichlet BC  in a certain sense, since it fixes the gauge fields on the boundary up to some trivial gauge transformations. To see this, let us recall the fact that $N^{\mu} {}^*\mathcal{F}_{\mu\nu}=0$ implies that $\gamma^{\mu}_{\rho}\gamma^{\nu}_{\sigma}\mathcal{F}_{\mu\nu}=F_{Q\rho\sigma}=0$ ($\gamma^{\mu}_{\rho}$ are projector operators from $N$ to $Q$), which fixes the field strength of induced gauge fields on the boundary.

For simplicity, we focus on the probe limit, i.e., AdS spacetime (\ref{AdSmetric}) with the bulk boundary $Q$
located at (\ref{Q}). This is sufficient for the study of the leading order of current. That is because the leading term of current (\ref{current1}) is of order $O(F)$, while the back-reaction of Maxwell fields to AdS is of order $O(F^2)$ \cite{Chu:2018ntx}.
For plane boundary,  $\mathcal{A}_{\mu}$ depends on
only the coordinates $z$ and $x$. And
the Maxwell's equations can be solved with
$\mathcal{A}_z=\mathcal{A}_z(z)$,
$\mathcal{A}_x=\mathcal{A}_x(x)$ and
\begin{eqnarray}\label{EOMvectorhigher}
  z \partial_x^2 \mathcal{A}_a -(d-3)\partial_z \mathcal{A}_a
  +z \partial_z^2 \mathcal{A}_a =0.
\end{eqnarray}
Similarly, the BCs (\ref{NBCa},\ref{DBCa}) become
\begin{eqnarray}
&&\text{Absolute BC} : ( \partial_x \mathcal{A}_a+\sinh\rho
  \partial_z \mathcal{A}_a ) \Big|_{x=-z \sinh \rho } =0,\label{NBCa1}\\
&&\text{Relative BC} : \ ( \partial_z \mathcal{A}_a-\sinh\rho
  \partial_x \mathcal{A}_a ) \Big|_{x=-z \sinh \rho } =0.\label{DBCa1}
\end{eqnarray}
Inspired by \cite{Miao:2017aba}, we consider the ansatz for the gauge field
\begin{eqnarray}\label{vectoransatz-5d}
\mathcal{A}_a=A^{(0)}_a+ A^{(1)}_a x f(\frac{z}{x}),
\end{eqnarray}
where $f(0)=1$ and $A^{(i)}_a$ are constants. The Maxwell's equations
(\ref{EOMvectorhigher}) become
\begin{eqnarray}
s (s^2+1) f''(s)-(d-3)f'(s)=0,
\end{eqnarray} 
which has the general solution
\begin{eqnarray}\label{vectorsolutionhigher}
  f(s)=1+\alpha_d  \frac{s^{d-2} \, _2F_1\left(\frac{d-3}{2},
    \frac{d-2}{2};\frac{d}{2};-s^2\right)}{d-2},
\end{eqnarray}
with $\alpha_d$ an integration constant. It should be mentioned that,
in order to get regular solutions at $x=0$, suitable analytic
continuation of the hypergeometric function should be taken when one
express the above solutions in terms of the coordinates $z$ and $x$.
Substituting (\ref{vectorsolutionhigher}) into (\ref{NBCa1},\ref{DBCa1}), we solve
the integration constant
\begin{equation}\label{integralconstantNBCa}
  \alpha_{A d} =\frac{(2-d) \text{csch}^3\rho (-\coth\rho)^d}{\text{csch}\rho \,
    _2F_1\left(\frac{d-3}{2},\frac{d-2}{2};\frac{d}{2};-\text{csch}^2\rho\right)
    (\coth \rho \ \text{csch}\rho)^d
    +(d-2) \cosh\rho \coth ^4\rho (-\text{csch}\rho)^d}.
\end{equation} 
for absolute BC and 
\begin{equation}\label{integralconstantDBCa}
  \alpha_{R d} =\frac{(2-d) (-\text{csch}\rho)^{2-d}}{\, _2F_1\left(\frac{d-3}{2},\frac{d-2}{2};\frac{d}{2};-\text{csch}^2\rho\right)}.
\end{equation}
for relative BC.
Similarly, suitable analytic
continuation of the hypergeometric function should be taken
in order to get smooth function
at $\rho=0$. For example, we have for $d=4,5$, 
\begin{eqnarray}\label{integralconstanta1}
 && \alpha_{A 4}=1,\qquad \ \ \ \ \ \ \  \ \ \  \ \ 
  \alpha_{A 5}=\frac{2}{\pi+4 \tan ^{-1}\left(\tanh \left(\frac{\rho }{2}
    \right)\right) },\\
&& \alpha_{R 4}=\frac{1}{1+\coth \rho},\qquad
  \alpha_{R 5}=\frac{2}{\pi+4 \tan ^{-1}\left(\tanh \left(\frac{\rho }{2}
    \right)\right)+2\text{csch} \rho }.\label{integralconstanta2}
\end{eqnarray}

Now we are ready to derive the holographic current for BCFT. Consider the variation of effective action for BCFT, in general, we have \cite{McAvity:1993ue}
\begin{eqnarray}\label{dIeff}
  \delta I_{\text{eff}}= \int_M\sqrt{g_0} J^i \delta A_{i}+\int_P\sqrt{h} [ j^{(0)a}\delta A^{(0)}_{a}+j^{(1)a}\delta A^{(1)}_{a}+...]
\end{eqnarray}
where $A^{(n)}_{a}$ are defined in the gauge field of BCFT $A_a=A^{(0)}_a+ A^{(1)}_a x +...$, $J^i $ and $j^{(n)a}$ are bulk and boundary operators respectively. In general, both $J^i $ and $j^{(n)a}$ are divergent. 

Near the boundary, there is no meaning to distinguish the divergent parts of the bulk current $J_i$ and boundary operators $j^{(n)}_i$. Only the combination of the bulk current and boundary operators have physical meaning. 
According to \cite{Lewkowycz:2014jia,Dong:2016wcf},
one can always regulate the effective action by excluding from its volume
integration a small strip of geodesic distance $\epsilon$ from the boundary.
Then there is no explicit boundary
divergences in this form of the effective action, however there are boundary
divergences implicit in the bulk effective action
which is integrated up to distance $\epsilon$.
Now the variation of effective action becomes
\begin{eqnarray} \label{key1}
\delta I_{\text{eff}}= \int_{M_{\epsilon}} \sqrt{g_0} \hat{J}^{i}\delta A_{i}
\end{eqnarray}
where $M_{\epsilon}$ denotes $x \ge \epsilon$. The boundary operators are absorbed into the total bulk current $\hat{J}^i$ in (\ref{key1}). One can recover the boundary operators after the integral up to $x=\epsilon$. Let us consider a simple example with $g_0=h=1$ and $\hat{J}^{i}=l^i/ x^m$. We have 
\begin{eqnarray} \label{exampleI}
\delta I_{\text{eff}}&=& \int_{x\ge \e}dx dy^{d-1}\frac{l^i}{x^m} \delta [A^{(0)}_i+ A^{(1)}_i x+... ]\nonumber\\
&=& \text{bulk terms}+\int_{x=\e}dy^{d-1}[\frac{l^i}{(m-1)\e^{m-1}} \delta A^{(0)}_i+ \frac{l^i}{(m-2)\e^{m-2}}  \delta A^{(1)}_i+... ],
\end{eqnarray}
which recovers the divergent parts of boundary operators.

Let us go on to consider the holographic theory. 
Consider the variation the on-shell action (\ref{action}) with respect to gauge fields, we have
\begin{eqnarray}\label{dI}
  \delta I= \int_M\sqrt{g} N_{\mu}\mathcal{F}^{\mu\nu}\delta \mathcal{A}_{\nu}+\int_Q\sqrt{\gamma} N_{\mu}\mathcal{F}^{\mu\nu}\delta \mathcal{A}_{\nu}
\end{eqnarray}
To make consistent (\ref{dI}) and (\ref{key1}), we should transform the integral on $Q$ into the integral on $M_{\e}$. There are two ways to do so. First, one can do the integral along the bulk direction $z$ on $Q$ and get integrals on $P$. Then we choose a suitable renormalization as above to transform the divergent terms on $P$ into the integral on $M_{\epsilon}$. Second, one can choose the same coordinates $(x,y_a)$ on $Q$ as on $M$ and rewrite $\delta \mathcal{A}_i|_Q$ in terms of $\delta A_i$. Then integral on $Q$ becomes integral on $M$. These two methods give the same results for the divergent parts of current. Below we take the second method for simplicity. The key point is that, on-shell, both $\mathcal{A}_{\mu}$ on $M$ and $Q$ are functions of background gauge fields $A_i$ of BCFT. For example, from (\ref{vectoransatz-5d}) we have on $M$
\begin{eqnarray}\label{AAM}
\mathcal{A}_i|_M=\lim_{z\to 0} \mathcal{A}_i= A^{(0)}_i+ \sum_{n=1}^{\infty}A^{(n)}_i x^n =A_i.
\end{eqnarray}
Similarly, from (\ref{Q},\ref{vectoransatz-5d}) we have on $Q$
\begin{eqnarray}\label{AAQ}
\mathcal{A}_i|_Q=A^{(0)}_a+ A^{(1)}_a x f(-\text{csch}\rho) +O(x^2)=A_i+ x \partial_x A_i \left( f(-\text{csch}\rho)-1 \right)+O(x^2)
\end{eqnarray}
Substituting (\ref{AAM}) and (\ref{AAQ}) into (\ref{dI}), we can derive the current via the definition 
\begin{eqnarray}\label{definecurrent}
J^i=\frac{1}{\sqrt{g_0}}\frac{\delta I}{\delta A_i},
\end{eqnarray}
where we have ignored the hat for the bulk current, $g_{0}$ is determinant of the metric of BCFT. For our case, we have $\sqrt{g_0}=1, \sqrt{g}=1/z^d, \sqrt{\gamma}=\cosh \rho \sinh^{d-1} \rho/x^d$. 

We discuss the current for absolute BC and relative BC separately. For absolute BC (\ref{NBCa}), the current on $Q$ vanishes and (\ref{dI}) is simplified as
\begin{eqnarray}\label{dI1}
  \delta I= \int_M\sqrt{g} N_{\mu}\mathcal{F}^{\mu\nu}\delta \mathcal{A}_{\nu}=\int_M \sqrt{g_0} \frac{\partial_z \mathcal{A}_i}{z^{d-3}} \delta A^i,
\end{eqnarray}
from which one can read off the current \cite{Chu:2018ntx}
\begin{eqnarray}\label{currentNBCa}
J_i=\lim_{z\to 0}\frac{\partial_z \mathcal{A}_i}{z^{d-3}}=
  -\alpha_{A d}\frac{F_{in} }{x^{d-3}} +O(\frac{1 }{x^{d-4}}).
\end{eqnarray}
In the above derivations, we have used (\ref{vectoransatz-5d},\ref{vectorsolutionhigher}) and $A^{(1)}_i= F_{ni}$. The near-boundary current (\ref{currentNBCa}) takes the expected form (\ref{current1}) with $\beta_d=-\alpha_{A d}$. Furthermore, it is proved in \cite{Chu:2018ntx} that the coefficients $\alpha_{A 4}, \alpha_{A 5}$ (\ref{integralconstanta1}) obey the universal relations (\ref{relationa1}).

Now let us go on to discuss the relative BC, which is one of the new results of this paper. For relative BC, there are non-trivial contributions to the current from the near-boundary region of $Q$. Substituting (\ref{DBCa1},\ref{vectoransatz-5d},\ref{AAQ}) into (\ref{dI}), we get
\begin{eqnarray}\label{dI2}
  \delta I&=&\int_M dx dy^{d-1}\frac{\alpha_{R d} F_{n}^{\ i} }{x^{d-3}} \delta A^i \nonumber\\
&+&\int_Q dx dy^{d-1} \frac{\coth ^2\rho (- \text{csch}\rho )^{3-d} f'(-\text{csch}\rho)F_{n}^{\ i}}{x^{d-3}} \delta \mathcal{A}_i|_Q, \nonumber\\
\end{eqnarray}
where $f(s), \alpha_{R d}$ are given by (\ref{vectorsolutionhigher},\ref{integralconstantDBCa}). Following the methods for absolute BC, one can derive the first line of (\ref{dI2}). As for the second line, we need to work out $\delta \mathcal{A}|_Q$, which can be derived from relative BC (\ref{DBCa}). As we have shown in the Introduction (discussions between (\ref{dIin}) and (\ref{DBCV2})), the relative BC (\ref{DBCa}) is equivalent to fixing the field strength of induced gauge fields
\begin{equation}\label{BCQ}
\mathcal{F}_{ij}|_Q= \partial_i\mathcal{A}_j|_Q-\partial_j\mathcal{A}_i|_Q=0,
\end{equation}
where $x^i=(x, y^a)$  are coordinates on the bulk boundary $Q$. Since we focus on the solutions depending on only $z$ and $x$ in the bulk, the induced gauge field $\mathcal{A}|_Q$ depends on only $x$:
\begin{eqnarray}\label{AQDBC0}
 \mathcal{A}|_Q= A^{(0)}+ \c_1 x +... + \c_m x^m+...
\end{eqnarray}
where $\c_m$ are constant vectors to be determined. Substituting $\mathcal{A}|_Q $ into (\ref{BCQ}), we solve $\gamma_i =0$ and thus
\begin{eqnarray}\label{AQDBC01}
 \mathcal{A}|_Q= A^{(0)}. 
\end{eqnarray}
From (\ref{AAM}) and (\ref{AQDBC01}), we have
\begin{eqnarray}\label{AQDBC1}
\delta \mathcal{A}_i|_Q= \delta A^{(0)}_i = \delta \left( A_i - x \partial_x A_i+..+ (-1)^m\frac{x^m}{m!} \partial^m_x A_i+...\right).
\end{eqnarray}
The cases for $d=4$ and $d>4$ are a little different. We discuss them respectively below.
For $d=4$, there are non-trivial contributions from $Q$ to the current. We have on leading term
\begin{eqnarray}\label{dI3}
  \delta I_4&=&\int_M dx dy^{3}\frac{\alpha_{R 4} F_{n}^{\ i} }{x} \delta A^i +\int_Q dx dy^{3} \frac{(1-\alpha_{R 4})F_{n}^{\ i}}{x} \delta[A_i+O(x)]\nonumber\\
&=& \int_M dx dy^{3}\frac{F_{n}^{\ i} }{x} \delta A^i,
\end{eqnarray}
which yields the current
\begin{eqnarray}\label{currentDBCa4d}
J_i=-\frac{F_{in} }{x} +O(1).
\end{eqnarray}
Note that (\ref{currentDBCa4d}) is exactly the same as the current with absolute BC (\ref{integralconstanta1},\ref{currentNBCa}). This is consistent with the fact that the current is independent of BCs in four dimensions. As we have mentioned above, the current with absolute BC  obey the universal law (\ref{relationa1}) \cite{Chu:2018ntx}. So does the current with relative BC (\ref{currentDBCa4d}). This is a non-trivial test of the validity of relative BC (Dirichlet BC) in holographic BCFT. 

As for $d\ge 5$, it is remarkable that the contributions on Q cancel out,
\begin{eqnarray}\label{dI4}
  \delta I_d&=&\int_M dx dy^{d-1}\frac{\alpha_{R d} F_{n}^{\ i} }{x^{d-3}} \delta A^i +\int_Q dx dy^{d-1} \frac{\alpha_{R d}F_{n}^{\ i}}{x^{d-3}} \delta[  \sum_{m=0}(-1)^m\frac{x^m}{m!} \partial^m_x A_i ]\nonumber\\
&=& \int_M dx dy^{d-1}\frac{\alpha_{R d}F_{n}^{\ i} }{x^{d-3}} \delta A^i.
\end{eqnarray}
As a result only the usual terms on $M$ contribute to the current. We thus have
\begin{eqnarray}\label{currentDBCa5d}
J_i=-\frac{\alpha_{R d} F_{in} }{x^{d-3}} +O(\frac{1}{x}),
\end{eqnarray}
which takes very similar expression as the current with absolute BC (\ref{currentNBCa}). In \cite{Chu:2018ntx}, we have proved for $d=5$ that the current with absolute BC (\ref{currentNBCa}) satisfies the universal relation (\ref{relationa1}). To do so, we take into account the back-reactions to AdS from the bulk Maxwell fields up to order $O(F^2)$ and then calculate the holographic Weyl anomaly. Finally, we compare the central charge $b_2$ in holographic Weyl anomaly with the current coefficient $\beta_5=-\alpha_{R 5}$ and verify (\ref{relationa1}). Following the same approach, we can prove that the current with relative BC also obey the universal relation (\ref{relationa1}). Since the proof is almost the same as that for absolute BC, we do not repeat it here.  In fact, one only needs to replace $\alpha_5$ by $\alpha_{R 5}$ in the proof of \cite{Chu:2018ntx}. 

Now we have proved that both absolute BC and relative BC can derive the correct one point function of current and universal relations between the current and Weyl anomaly. This is a strong support that both absolute BC (Neumann BC) and relative BC (Dirichlet BC) are consistent in holographic BCFT.

\subsection{Exact solutions}

In the above subsection, we have investigated perturbation solutions in holographic BCFT. In general, it is a non-trivial problem to find exact solutions to Einstein-Maxwell equations after imposing BCs on $Q$. Here we notice that the four-dimensional Reissner-Nordstr$\ddot{\text{o}}$m black holes are exact solutions to holographic BCFT, provided that the bulk boundary $Q$ is perpendicular to the AdS boundary $M$ \footnote{It should mentioned that the authors of \cite{Chang:2018pnb} also notice that Schwarzschild-AdS black holes are solutions to holographic BCFT when the $Q$ is perpendicular to the AdS boundary $M$.}. Interestingly, the magnetic charged Reissner-Nordstr$\ddot{\text{o}}$m black holes satisfy the relative (magnetic) BC and predict that there is a constant boundary current on $P$, when a constant magnetic field is applied on $M$ for 3d BCFT.

Now let us list the main results. We find the electric charged Reissner-Nordstr$\ddot{\text{o}}$m black holes satisfy the Neumann BC (\ref{NBCg}) and  absolute BC (\ref{NBCa}) imposed on $Q$.
\begin{eqnarray}\label{ExactSolutionI}
&&\text{metric}: \ \ \ \ \ ds^2=\frac{dz^2/f(z)-f(z)dt^2+dx^2+dy^2}{z^2}, \nonumber\\
&&\text{gauge field}: \mathcal{A}=q z dt,  \\
&&\text{Q}: \ \ \ \ \ \ \ \  \ \ \ \  x=0, \nonumber
\end{eqnarray}
where $f(z)=1-M z^3+q^2 z^4/4$ and recall that we have $T=\rho=0$. In the dual picture, we have zero background gauge field
\begin{eqnarray}\label{backgroundAI}
A=\lim_{z\to 0}\mathcal{A}=0,
\end{eqnarray}
but non-zero charge density on $M$
\begin{eqnarray}\label{chargedensityI}
J_t=q. 
\end{eqnarray}
The above picture is a little trivial, since it does not tell us new stories compared with the case without boundary. 

Now let us consider magnetic charged Reissner-Nordstr$\ddot{\text{o}}$m black holes, which is more interesting. One can check that, for $T=\rho=0$, the magnetic charged Reissner-Nordstr$\ddot{\text{o}}$m black holes satisfy the Neumann BC (\ref{NBCg}) for metrics and the relative BC (\ref{DBCa}) for gauge fields
\begin{eqnarray}\label{ExactSolutionII}
&&\text{metric}: \ \ \ \ \ ds^2=\frac{dz^2/f(z)-f(z)dt^2+dx^2+dy^2}{z^2}, \nonumber\\
&&\text{gauge field}: \mathcal{A}=(a_y+B x )dy,  \\
&&\text{Q}: \ \ \ \ \ \ \ \  \ \ \ \  x=0, \nonumber
\end{eqnarray}
where $f(z)=1-M z^3+B^2 z^4/4$ and the constant $a_y$ is the background gauge field on the boundary $P$. In the dual picture, we have a constant magnetic field on $M$
\begin{eqnarray}\label{backgroundAII}
B=F_{xy},
\end{eqnarray}
and non-zero current on $P$
\begin{eqnarray}\label{boundarycurrentII}
j_y=B z_h,
\end{eqnarray}
where $z_h$ is the location of outer horizon, i.e., $f(z_h)=1-M z_h^3+B^2 z_h^4/4=0$. To derive the boundary current (\ref{boundarycurrentII}), let us consider the variation of action. Substituting (\ref{ExactSolutionII}) into (\ref{dI}), we have
\begin{eqnarray}\label{dI5}
  \delta I=\int_Q  dz dt dy\  B \ \delta a_y= \int_P dt dy\  B z_h \ \delta a_y, 
\end{eqnarray}
which gives the boundary current (\ref{boundarycurrentII}). Notice that since the boundary current is finite, there is no need to perform the approach (\ref{key1}) of sect.2.2, which is developed for the divergent parts of current. 

Let us make some comments on the boundary current (\ref{boundarycurrentII}). First of all, the holographic BCFTs with relative BCs predict that there is a constant boundary current on $P$, when a constant magnetic field is applied on $M$ in three dimensions.
Secondly, the current depends on both the magnetic field and the temperature
\begin{eqnarray}\label{tem}
 T_{\text{tem}}=-\frac{f'(r_h)}{4\pi}=z_h^2(3M-B^2 z_h),
\end{eqnarray}
where we have $M^4 \ge (4/27) B^6$ in order to avoid the naked singularity, and the temperature vanishes when the lower bound of mass is saturated.
Thirdly, the absolute value of current decreases as the temperature increases.
In particular, the current gains its maximum absolute value at zero temperature
\begin{eqnarray}\label{boundarycurrentlowT}
j_y=12^{\frac{1}{4}}\frac{B}{\sqrt{|B|}},
\end{eqnarray}
while vanishes in high temperature limit
\begin{eqnarray}\label{boundarycurrenthighT}
j_y= \frac{3B}{T_{\text{tem}}}+O(\frac{1}{T_{\text{tem}}^2}).
\end{eqnarray}
Fourthly, it should be mentioned that $T=\rho=0$ is the dual BC preserving the maximum (half) supersymmetry \cite{Astaneh:2017ghi}. And it seems that the non-renormalization theorem holds for such BCs. As a result,  the dual strongly-coupled BCFTs can be mimicked by free BCFTs. Thus the above prediction would also apply to some kinds of free BCFTs with suitable BCs. For general BCs, there would be both bulk current and boundary current. Last but not least, it is expected that a constant magnetic field in the bulk $M$ can induce constant boundary current on the boundary $P$ for general boundary quantum field theory (BQFT). The boundary current increases as the magnetic field is enhanced, while decreases as the temperature and the mass of charged particles increase. In the zero temperature and mass limit, from the dimension analysis and isotropy, the boundary current takes the general form
\begin{eqnarray}\label{boundarycurrentBQFT}
j_y=\lambda_0\ c\sqrt{\frac{e^3}{\hbar}}\frac{B}{\sqrt{|B|}},
\end{eqnarray}
where we have recoverd the units, $e$ is the charge, $c$ is the speed of light, $\hbar$ is the Planck constant and $\lambda_0$ is a dimensionless constant determined by the theory.  It is remarkable that this current is detectable. For $B=1T$ and $\gamma_0=1$, we have
\begin{eqnarray}\label{boundarycurrentBQFTlab}
j_y\approx 0.0003 A.
\end{eqnarray}
It is interesting to measure such boundary current in some (1+2) dimensional systems with boundaries, such as graphene \cite{graphene}. To enhance this effect, one should try to decrease the temperature and the effective mass of charged particles in materials. 

To end this section, we remark that the holographic BCFT (\ref{action}) has electromagnetic duality in four dimensions. And the absolute BC and relative BC are dual to each other.  This is another support that both absolute BC and relative BC  are well-defined. 
Let us briefly discuss the electromagnetic duality in four dimensions below.  By `electromagnetic duality', it means the theory is invariant under the following transformations
\begin{eqnarray}\label{EBduality}
\mathcal{F}_{\mu\nu}\to {}^*\mathcal{F}_{\mu\nu}=\frac{1}{2}\epsilon_{\mu\nu\rho\sigma} \mathcal{F}^{\rho\sigma},
\end{eqnarray}
which transform electric field and magnetic field into each other, i.e., $(E, B) \to (c B, -E/c)$ (c is  the velocity of light). Applying the formula 
\begin{eqnarray}\label{EBdualityformula}
\mathcal{F}_{\mu\nu} \mathcal{F}^{\mu\nu} = {}^*\mathcal{F}_{\mu\nu}{}^*\mathcal{F}^{\mu\nu},
\end{eqnarray}
we find that the action of holographic BCFT (\ref{action})  is indeed invariant under the electromagnetic duality transformations (\ref{EBduality}).  And it is obvious that  (\ref{EBduality}) transform absolute BC $N^{\mu}\mathcal{F}_{\mu\nu}=0$ (\ref{NBCa}) into relative BC $N^{\mu}{}^*\mathcal{F}_{\mu\nu}=0$ (\ref{DBCa}). So the absolute BC and relative BC are indeed dual to each other. 
Similarly, one can show that the electric charged black hole (\ref{ExactSolutionI}) and magnetic charged black holes (\ref{ExactSolutionII}) are dual to each other, provided that we set the parameter $B=q$.  First, it is obvious from (\ref{ExactSolutionI},\ref{ExactSolutionI}) that the metrics of electric charged black hole and magnetic charged black holes are exactly the same for $B=q$. Second, one can show that the electromagnetic duality transformations (\ref{EBduality}) transform the field strengths of electric charged black hole into the field strengths of magnetic charged black hole.   From (\ref{ExactSolutionI}), we get the non-zero components of field strength $\mathcal{F}_{zt}=q$ for electric charged black hole. Under the electromagnetic duality transformations (\ref{EBduality}), it transforms to
\begin{eqnarray}\label{FztFxy}
\mathcal{F}_{zt}\to {}^*\mathcal{F}_{zt}=\mathcal{F}_{xy}=q,
\end{eqnarray}
which is exactly the field strength (\ref{backgroundAII}) of magnetic charged black hole for $q=B$. Now we finish the proof of our statements. 
To summary, the  electromagnetic duality transformations (\ref{EBduality}) transform electric charged black hole satisfying absolute BC into  magnetic charged black hole satisfying relative BC.

\section{Holographic BCFT with Dirichlet BC}

In this section, we study the Dirichlet BC for holographic BCFT. We find that, similar to Neumann BC, Dirichlet BC can also yield the expected one point function of stress tensors and the universal relations between stress tensors and Weyl anomaly. 

Consider the variations of action (\ref{action}) with respect to the metric, we have
\begin{eqnarray}\label{dIDBC}
  \delta I=\int_Q \sqrt{\gamma} [(K-T) \gamma^{ij}-K^{ij}]\delta \gamma_{ij}.
\end{eqnarray} 
To have a well-defined action principle, one can impose either Neumann BC (\ref{NBCg}) or the Dirichlet BC
\begin{eqnarray}\label{DBCg}
\delta \gamma_{ij} |_Q=0.
\end{eqnarray} 
Dirichlet BC fixes the induced metric on the boundary. To proceed, we must specify what kinds of metrics we choose. The most simple and natural one is the AdS metric, which satisfes
\begin{eqnarray}\label{DBCAdS}
R^Q_{ijkl}+\text{sech}^2 \rho (\gamma_{ik}\gamma_{jl}-\gamma_{il}\gamma_{jk})=0,
\end{eqnarray} 
with $\rho$ a free parameter of the model. The Dirichlet BC (\ref{DBCAdS}) is the central assumption of this paper.  Of course, one can choose the other metrics such as those of asymptotically AdS, black holes and so on. For simplicity, we focus on AdS and leave the study of other choices to future work. Another reason to choose the AdS metric is that it can be easily written into covariant form (\ref{DBCAdS}). We remark that  (\ref{DBCAdS}) is the gravitational counterpart of the relative BC (\ref{DBCa}) for gauge fields, since they both fix the curvatures of induced fields on the boundary.

The good BCs for holographic BCFT should satisfy the following requirements:
\begin{itemize}
\item[(1)] 
{\it 
It should be neither too strong nor too weak. In particular, it allows AdS to be a solution. }

If the BC allows no solutions or only a limited number of solutions, it is too strong. On the other hand, it is too weak if all the solutions to EOM are allowed. In general, there is a narrow window of consistent BCs. We also hope AdS is a solution so that we can apply AdS/CFT.   

\item[(2)] 
{\it 
It can fix the location of bulk boundary $Q$.}

This is the central problem in the construction of AdS/BCFT. 

\item[(3)] 
{\it 
It can produce the expected one point function of stress tensor (current), Weyl anomaly and universal relations between them.}

\end{itemize}

In this section, we will show that Dirichlet BC satisfies all the above requirements. So the holographic BCFT with Dirichlet BC is well-defined. 
As a quick check, we notice that the Poincare AdS
(\ref{AdSmetric}) together with the embedding function of $Q$ (\ref{Q}) are indeed solutions to the Dirichlet BC (\ref{DBCAdS}). As a result, holographic BCFT with Dirichlet BC share most of the advantages of holographic BCFT with Neumann BC \cite{Takayanagi:2011zk}. In particular, it obeys the holographic g-theorem due to the fact that the two kinds of holographic BCFT have the same g-functions
 To see this, let us show some details below. We require that the Poincare AdS
(\ref{AdSmetric}) is a solution to Dirichlet BC (\ref{DBCAdS}) for some suitable embedding function of $Q$. For simplicity, we assume the embedding function is independent of $y_a$
\begin{eqnarray}\label{DBCAdSQ}
x=-F(z).
\end{eqnarray}
Substituting (\ref{DBCAdSQ}) into (\ref{AdSmetric}), we get the induced metric on $Q$
\begin{eqnarray}\label{inducedAdSQ}
ds_Q^2=\frac{(1+F'(z)^2)dz^2+\delta_{ab}dy^ady^b}{z^2}. 
\end{eqnarray}
Imposing the DBC (\ref{DBCAdS}), we can solve $F(z)$. From $(z, y_1,z, y_1)$ components of  DBC (\ref{DBCAdS}), we get
\begin{eqnarray}\label{E1212}
1+F'(z)^2-\text{sech}^2\rho \left(1+F'(z)^2\right)^2+\frac{1}{2}z \left(F'(z)^2\right)'=0.
\end{eqnarray}
Since the curvatures of (\ref{DBCAdS}) have only one independent component in two dimensions, there is only one independent BC (\ref{E1212}) for $d=2$.   As for $d\ge 3$, we have more independent components of curvatures and thus we have more BCs in addition to (\ref{E1212}).
For example, from $(y_1,y_2,y_1,y_2)$ components of  DBC (\ref{DBCAdS}), we derive
\begin{eqnarray}\label{DBCAdSQsolution1}
F'(z)^2=\sinh^2 \rho.
\end{eqnarray}
It is clear that there is no $(y_1,y_2,y_1,y_2)$ component for DBC (\ref{DBCAdS}) in two dimensions.  Thus (\ref{E1212}) works for both $d=2$ and $d\ge 3$, but (\ref{DBCAdSQsolution1}) works only for $d\ge 3$. 

Let us firstly discuss the case for $d\ge 3$. One can check that if (\ref{DBCAdSQsolution1}) is satisfied, all the components of (\ref{DBCAdS}) are automatically satisfied. As a quick check, we find this is indeed the case for (\ref{E1212}), i.e., the $(z, y_1,z, y_1)$ components of  DBC (\ref{DBCAdS}). Thus let us focus on (\ref{DBCAdSQsolution1}). 
We choose the boundary at $x=0$ which fixes the integration constant $F(0)=0$. We make the choice that 
\begin{eqnarray}\label{DBCAdSQchoice}
F'(z)=\sinh\rho.
\end{eqnarray}
in order to get same solution as Neumann BC (\ref{Q}). Note that, the trace of extrinsic curvature is positive for this choice for $\rho>0$.  Finally we obtain the embedding function (\ref{Q}). 

Now let us turn to the case for $d=2$.  As we have mentioned before, there is only one independent BC (\ref{E1212})  for $d=2$. Solving (\ref{E1212}), we obtain
\begin{eqnarray}\label{DBCAdSQsolution2}
F'(z)^2=\sinh^2\rho-\frac{z^2 \cosh ^2\rho}{2 e^{c_1 \cosh ^2\rho}+z^2},
\end{eqnarray}
where $c_1$ is a integration constant. 
Note that the left hand side of (\ref{DBCAdSQsolution2}) is positive, while the right hand side of (\ref{DBCAdSQsolution2}) could be negative for sufficiently large $z$. To avoid inconsistency, we must have $c_1=\infty$ and (\ref{DBCAdSQsolution2}) reduces to (\ref{DBCAdSQsolution1}). Following the above method, we get the embedding function (\ref{Q}) for $d=2$. 

Now we have shown that, holographic BCFTs with DBC and NBC have the same solutions, i.e., 
Poincare AdS (\ref{AdSmetric}) together with the embedding function of $Q$ (\ref{Q}). As a result, they have the same boundary entropy and boundary central charges related to Euler densities and thus both obey g-theorem. Let us show more details below.

For simplicity, we focus on the holographic g-theorem for 2d BCFTs.  Since holographic BCFTs with NBC and DBC have the same solutions, the following discussions apply to both of them. The g-theorem claim that the boundary degree of freedom (g-function) decrease under the RG flow.  On the AdS boundary $z=0$, we require that the g-function $g(z)$ reduce to boundary entropy, which is a natural candidate for  boundary degree of freedom. According to \cite{Takayanagi:2011zk}, the boundary entropy is defined by the difference between entanglement entropy with $\rho=\rho^*$ and that with $\rho=0$
\begin{equation}\label{boundaryentropy0}
S_{\text{bdy}}=S_{A}(\rho^*)-S_{A}(0), 
\end{equation}
where the subregion $A$ is defined by $0\ll x \ll l$. One can use RT formula \cite{Ryu:2006bv} to calculate the holographic entanglement entropy above. Following exactly \cite{Takayanagi:2011zk}, we have
\begin{equation}\label{boundaryentropy}
S_{\text{bdy}}=\frac{\rho^*}{4}.
\end{equation}
 Without loss of generality, we can choose g-function as
\begin{equation}\label{gfunction}
g(z)= -x'(z)
\end{equation}
so that we have $g(0)= \sinh \rho^* = \sinh (4 S_{\text{bdy}})$ from (\ref{Q}). 

Now imposing the null energy condition on $Q$ \cite{Takayanagi:2011zk}, we get
\begin{equation}\label{NEC}
(K_{\mu\nu}-K\gamma_{\mu\nu})m^{\mu} m^{\nu}= \frac{x''(z)}{z[1+x'(z)^2]^{\frac{3}{2}}} \ge 0,
\end{equation}
where $T_{\mu\nu}=(K_{\mu\nu}-K\gamma_{\mu\nu})$ are the Brown-York stress tensor on $Q$ and the null vectors are given by $m^{\mu}=(m^z, m^t, m^x)=\left((1+x'^2)^{-1/2},\pm 1, -x'(1+x'^2)^{-1/2}\right)$.  From (\ref{gfunction},\ref{NEC}), we derive
\begin{equation}\label{gtheorem}
\partial_ z g(z) \le 0,
\end{equation}
which shows that the g-function (boundary degree of freedom) is a monotonically decreasing function under the RG flow. Note that $x(z)$ in our notation differs by a minus sign from that of  \cite{Takayanagi:2011zk}. Recall that we have choose $K>0$ for the solutions. If we choose $K<0$ instead, we would get $S_{\text{bdy}}=-\frac{\rho}{4}$ and thus negative g-function, which disagrees with g-theorem.

As a summary, the Dirichlet BC (\ref{DBCAdS}) indeed allows AdS to be a solution. 
It is remarkable that, in addition to the Dirichlet BC (\ref{DBCAdS}), we impose an extra condition that the extrinsic curvature is positive definite $K\ge 0$ for $\rho\ge 0$ \footnote{We focus on the case $\rho\ge 0$ for reasons which will be made clear in sect.4.}. This requirement is reasonable. That is because, according to \cite{Witten:2018lgb}, the extrinsic curvature should be either positive definite or negative definite in order to have a well-defined quantization for Dirichlet BC. We make the physical choice $K\ge 0$ in order to satisfy the holographic g-theorem.

\subsection{Casimir effects and Weyl anomaly}

Let us quickly review the Casimir effects and Weyl anomaly for BCFTs. It is found in \cite{Deutsch:1978sc} that
the renormalized stress tensor of BCFT is divergent
near the boundary,
\begin{eqnarray}\label{stress0}
\la T_{ij} \ra  =-2\bar{\a}_d\frac{ \bar{k}_{ij}}{x^{d-1}}, \quad  x \sim 0,
\end{eqnarray}
where $x$ is the proper distance from the boundary, $\bar{k}_{ij}$ is the traceless part of extrinsic curvature and
$\alpha$ is a constant which depends only on the kind of BCFT under consideration.  The coefficient $\a$ fixes the shape
dependence of the leading Casimir effects of BCFT. 

Interestingly, the authors of \cite{Miao:2017aba} observe that the Casimir coefficient is related to the central charge of Weyl anomaly. For example,  there are universal relations 
\begin{eqnarray}\label{relationg}
\bar{\a}_3=b_2 ,\  \ \bar{\a}_4=-\frac{b_4}{2},
\end{eqnarray}
where $b_i$ are boundary central charges of Weyl anomaly of 3d BCFT and 4d BCFT \cite{Jensen:2015swa,Fursaev:2015wpa, Herzog:2015ioa}, respectively
\begin{eqnarray}\label{3dBWA}
 \mathcal{A} =\int_P\sqrt{h}(b_1 \mathcal{R}+b_2 \text{Tr}
  \bar{k}^2), 
\end{eqnarray}
\begin{eqnarray}\label{4dBWA}
  \mathcal{A}=\text{Bulk Weyl anomaly}+\int_{P}\sqrt{h}( b_3 \text{Tr}
  \bar{k}^3 + b_4 C^{ac}_{\ \ \ b c}
  \bar{k}_{\ a}^b).\;\;
\end{eqnarray}
Under some assumptions, the authors of \cite{Miao:2017aba} further check that the holographic BCFT with Neumann BC \cite{Takayanagi:2011zk} is consistent with the universal relations (\ref{relationg}) between Casimir effects and Weyl anomaly. In this paper, we give a solid proof of the universal relations (\ref{relationg}) for both the holographic BCFTs with Dirichlet BC (\ref{DBCAdS}) and Neumann BC \cite{Takayanagi:2011zk}.

\subsection{Solutions and Stress Tensors}

To study the one point function of stress tensor, one needs to derive the perturbation solutions around
(\ref{AdSmetric}) and (\ref{Q}). Following \cite{Miao:2017aba}, we take the following ansatz for the metric
\begin{eqnarray}\label{bulkmetric}
&& ds^2=\frac{1}{z^2}\Big{[} dz^2+dx^2 
+\big(\delta_{ab}-2x \bar{k}_{ab} f(\frac{z}{x})-2x \frac{k}{d-1}\delta_{ab}
 \big)dy^a dy^b+O(k^2)\Big{]},
\end{eqnarray}
and the embeding function of $Q$
\begin{eqnarray}\label{Q1}
x=-\sinh\rho \ z + \lambda_2 k z^2+ O(k^2) 
\end{eqnarray}
where $k$ denotes the trace of the extrinsic curvature, $f(s)$ is a function and $\lambda_2$ is a constant to be determined.  We
require that
\be \label{fXQ}
f(0)=1
\ee
so
that the metric of BCFT takes the form in Gauss
normal coordinates
\be \label{GNC}
ds^2=dx^2+ \left(\delta_{ab}-2x k_{ab}+O(\partial^2)\right) dy^ady^b.
\ee
Note that we expand the solutions in extrinsic curvatures, or equivalently, the derivatives $O(k)\sim O(\partial)$. Besides, we focus on the case without $y$ dependent, which is sufficient to obtain the Casimir coefficients and central charges of Weyl anomaly. 

Substituting (\ref{bulkmetric}) into Einstein equations,
we obtain at the order $O(k)$ a single equation
\begin{eqnarray}\label{EOM}
s \left(s^2+1\right) f''(s)-(d-1) f'(s)=0,
\end{eqnarray}
which can be solved as 
\begin{eqnarray}\label{solutiong}
f(s)=1+\bar{\a}_d \frac{s^d \,
 {}_2F_1\left(\frac{d-1}{2},\frac{d}{2};\frac{d+2}{2};-s^2\right)}{d}.
\end{eqnarray}
Imposing either Neumann BC \cite{Takayanagi:2011zk} or Dirichlet BC (\ref{DBCAdS}), we can fix the location of $Q$ (\ref{Q1}) and derive
\begin{eqnarray}\label{Qlambda2}
\lambda_2=\frac{\cosh ^2\rho}{2 (d-1)}.
\end{eqnarray}
In fact, $\lambda_2$ can be fixed by the symmetry of asymptotically AdS \cite{Chu:2017aab}. Thus it is universal and independent of BCs. Now the BCs on $Q$ (\ref{Q1},\ref{Qlambda2}) become
\begin{eqnarray}
&&\text{Neumann BC} : \coth\rho f'(-\text{csch}\rho)+\text{sech}\rho  f(-\text{csch}\rho)=0,\label{NBCg1}\\
&&\text{Dirichlet BC} : \ f(-\text{csch}\rho) =0.\label{DBCg1}
\end{eqnarray}
Substituting (\ref{solutiong}) into the above BCs, we obtain the integration constant for Neumann BC and Dirichlet BC respectively
\begin{eqnarray}
&&\bar{\alpha}_{Nd}=\frac{-d \cosh ^d\rho}{(-\coth\rho)^d \, _2F_1\left(\frac{d-1}{2},\frac{d}{2};\frac{d+2}{2};-\text{csch}^2\rho \right)+d \cosh ^2\rho \coth\rho},\label{aN}\\
&&\bar{\alpha}_{Dd}=\frac{-d (-\text{csch}\rho)^{-d}}{\, _2F_1\left(\frac{d-1}{2},\frac{d}{2};\frac{d+2}{2};-\text{csch}^2\rho \right)}.\label{aD}
\end{eqnarray}
We remark that the gravitational solutions (\ref{solutiong},\ref{aN},\ref{aD}) in $D=(d+1)$ dimensions are exactly the same as the solutions of gauge fields  (\ref{vectorsolutionhigher},\ref{integralconstantNBCa},\ref{integralconstantDBCa}) in $(D+2)$ dimensions. As we have mentioned before,
suitable analytic
continuation of the hypergeometric function should be taken
in order to get smooth function
at $\rho=0$. For example, we have for $d=4,5$, 
\begin{eqnarray}\label{integralconstantg1}
 && 
  \bar{\alpha}_{N 3}=\frac{2}{\pi+4 \tan ^{-1}\left(\tanh \left(\frac{\rho }{2}
    \right)\right) }, \ \ \ \ \ \ \ \ \ \ \ \  \bar{\alpha}_{N 4}=\frac{1}{2(1+\tanh \rho) },\\
&&
  \bar{\alpha}_{D 3}=\frac{2}{\pi+4 \tan ^{-1}\left(\tanh \left(\frac{\rho }{2}
    \right)\right)+2\text{csch} \rho }, \ \bar{\alpha}_{D 4}=\frac{\tanh \rho}{(1+\tanh \rho)^2 }. \label{integralconstantg2}
\end{eqnarray}

Now let us go on to derive the holographic stress tensors. We firstly discuss the case of Neumann BC. Using
\eq{bulkmetric}, \eq{solutiong}, we can derive the holographic stress
tensor \cite{deHaro:2000vlm}
\begin{eqnarray}\label{NBCholoTij}
T_{ij} =d h^{(d)}_{ij}=-2 \bar{\a}_{N d} \frac{
  \bar{k}_{ij} }{x^{d-1}} +O(k^2),
\end{eqnarray}
which takes the expected form
\eq{stress0}. Here  $ h^{(d)}_{ij}$ is defined in the Fefferman-Graham expansion of the
asymptotic AdS metric
\begin{eqnarray}\label{hnCDFGAdS}
ds^2=\frac{dz^2}{z^2}+\frac{1}{z^2}( g^{(0)}_{ij} + z^2  g^{(1)}_{ij}+\cdots+z^d  h^{(d)}_{ij}+\cdots   )dy^idy^j.
\end{eqnarray}
Now let us turn to the case of Dirichlet BC. Similar to the current of gauge fields, in general, there are potential contributions from the bulk boundary $Q$ to the stress tensors. Taking the metric variation of the action (\ref{action}) with $T=(d-2) \tanh \rho$ and using (\ref{Q1},\ref{Qlambda2},\ref{solutiong},\ref{DBCg1}), we have at leading order
\begin{eqnarray}\label{dIDBC}
  \delta I&=&\int_M dx dy^{d-1} (-\bar{\a}_{D d} \frac{
  \bar{k}^{ij} }{x^{d-1}}) \delta g^{(0)}_{ij}\nonumber\\
&+&\int_Q dx dy^{d-1}\frac{\coth^2 \rho (-\sinh \rho)^{d-3} f'(-\text{csch}\rho) \bar{k}^{ij}  }{x^{d-3}} \delta \gamma_{ij},
\end{eqnarray}
where $\gamma_{ij}$ is the AdS metric due to Dirichlet BC (\ref{DBCAdS}). Recall that we focus on the case that $\gamma_{ij}$ is a function of only $x$. Then the most general form of $\gamma_{ij}$ takes 
\begin{eqnarray}\label{gammaij}
ds^2= \gamma_{ij}dx^idx^j=\frac{\cosh^2 \rho\ f^2_1(x, k,q,...) dx^2+ \sinh^2 \rho\  f^2_2(x, k,q,...) h_{ab} dy^ady^b}{x^2},
\end{eqnarray}
where $f_1=1-x\frac{f'_2}{f_2}$, $h_{ab}=\delta_{ab}$ and $(h_{ab}, k_{ab}, q_{ab},...)$ are defined in the Gauss normal coordinate
\begin{eqnarray}\label{GNC1}
ds^2=g^{(0)}_{ij}dx^idx^j=dx^2+(h_{ab}-2x k_{ab}+x^2 q_{ab}+...)dy^ady^b.
\end{eqnarray}
To derive the divergent parts of stress tensor via $T^{ij}=\frac{2 \delta I}{\sqrt{g^{(0)}}\delta g^{(0)}_{ij}}$, 
we can replace $(h_{ab},k_{ab}, q_{ab},...)$ in $\gamma_{ij}$ by
\begin{eqnarray}\label{hkq1}
&&h_{ab}=\sum_{m=0}(-1)^m \frac{x^m}{m!}\partial^m_x g^{(0)}_{ab} ,\nonumber\\
&&k_{ab}=-\frac{1}{2}\sum_{m=0}(-1)^m \frac{x^m}{m!}\partial^{m+1}_x g^{(0)}_{ab},\\
&&q_{ab}=\frac{1}{2} \sum_{m=0}(-1)^m \frac{x^m}{m!}\partial^{m+2}_x g^{(0)}_{ab}.\nonumber
\end{eqnarray}
In this way, we derive 
\begin{eqnarray}\label{dgammaijDBC4d}
 \delta \gamma_{ab}=\frac{\sinh^2 \rho}{x^2} \left(f^2_2 \sum_{m=0} \frac{(-x)^m}{m!}\partial^m_x \delta g^{(0)}_{ab}+  h_{ab}\delta f^2_2 \right).
\end{eqnarray}
Note that $\delta \gamma_{xx}$ is not important for our purpose, since the stress tensors on $Q$ (\ref{dIDBC}) have no $xx-$components at the leading order. 
Substituting (\ref{dgammaijDBC4d}) into (\ref{dIDBC}), we derive 
\begin{eqnarray}\label{dIDBC1}
\delta I&=&\int_M dx dy^{d-1} (-\bar{\a}_{D d} \frac{
  \bar{k}^{ij} }{x^{d-1}}) \delta g^{(0)}_{ij}\nonumber\\
&+&\int_Q dx dy^{d-1}\frac{f_3(\rho)\bar{k}^{ab}  }{x^{d-1}} \left(f^2_2 \sum_{m=0} \frac{(-x)^m}{m!}\partial^m_x \delta g^{(0)}_{ab}+  h_{ab}\delta f^2_2 \right)\nonumber\\
&=&\int_M dx dy^{d-1} (-\bar{\a}_{D d} \frac{
  \bar{k}^{ij} }{x^{d-1}}) \delta g^{(0)}_{ij}
\end{eqnarray}
Similar to the case of gauge fields, the contributions on $Q$ vanish. Note that the stress tensor with $d=2$ is the counterpart of the current with $d=4$ of sect.2.2. Similar to the case of current, there would be non-trivial contributions on $Q$ to the stress tensor with $d=2$. However, due to  $\bar{k}^{ab}=0$ for $d=2$, such potential  contributions vanish. 
From (\ref{dIDBC1}) we finally obtain the stress tensor for Dirichlet BC 
\begin{eqnarray}\label{DBCholoTij}
T_{ij} =-2 \bar{\a}_{D d} \frac{
  \bar{k}_{ij} }{x^{d-1}} +O(k^2),
\end{eqnarray}
which takes the same form as Neumann BC (\ref{NBCholoTij}).  In the next section, we shall test the universal relations (\ref{relationg})  for both Dirichlet BC and Neumann BC.

\subsection{Holographic Weyl anomaly}

In this section, we derive the holographic Weyl anomaly and verify the universal relations (\ref{relationg}) between Casimir effects and Weyl anomaly for 3d BCFT and 4d BCFT. To do so, we need to work out the perturbation solutions up to order $O(k^2)\sim O(\partial^2)$ for 3d BCFT and $O(k^3, k q)\sim O(\partial^3)$ for 4d BCFT, respectively. Since the calculations are quite complicated, below we take 3d BCFT as an example to illustrate the key points of derivations and only claim the final result for 4d BCFT. 

Following \cite{Miao:2017aba}, we take the following ansatz for metrics
\begin{eqnarray}\label{bulkmetrick2}
&& ds^2=\frac{1}{z^2}\Big{[} dz^2+ \left(1+x^2
  X_2(\frac{z}{x})+x^3
  X_3(\frac{z}{x})+...\right)dx^2  \nonumber \\
&& +\left(\delta_{ab}-2x \bar{k}_{ab} f(\frac{z}{x})-2x \frac{k}{d-1}
  \delta_{ab}  + x^2 Q_{ab}(\frac{z}{x})+ x^3 H_{ab}(\frac{z}{x}) +...\right)dy^a dy^b\Big{]}  \nonumber\\
\end{eqnarray}
where
the functions $ X_2(\frac{z}{x})$ and $ Q_{ab}(\frac{z}{x})$ are of
order $O(\partial^2)$, $ X_3(\frac{z}{x})$ and $ H_{ab}(\frac{z}{x})$ are of
order $O(\partial^3)$ and $...$ denotes higher orders. We 
set that
\be \label{fXQ}
f(0)=1,\quad X_2(0)=X_3(0)=0, \quad Q_{ab}(0)=q_{ab},  \quad H_{ab}(0)=h_{ab}
\ee
so
that the metric of BCFT takes the form in Gauss normal coordinates
\be \label{GNC3}
ds_M^2=dx^2+ \left(\delta_{ab}-2x k_{ab}+x^2q_{ab}+x^3 h_{ab}+...\right) dy^ady^b.
\ee
For simplicity, we focus on the solutions without $y_a$ dependence. We further
set $k_{ab}=\text{diag}(k_1,..., k_{d-1}),   q_{ab}=\text{diag}(q_1,..., q_{d-1})$,
where $k_a, q_a$ are constants. Then the embedding function of $Q$ takes the form
\be \label{Qhigherorder}
x=-\sinh\rho \ z + \frac{\cosh ^2\rho}{2 (d-1)} k z^2+ c_3 z^3+ ...
\ee
where the second term is fixed by the symmetry of asymptotically AdS \cite{Chu:2017aab}, $c_i$ are constants of order $O(\partial^{i-1})$ which can be determined by BCs. 

Now let us focus on the second order solutions for 3d BCFT. Substituting (\ref{bulkmetrick2}) into the
Einstein equations, we solve \cite{Miao:2017aba}
 \bea \label{Qabapp1}
&&f(s) =1-\a_1 (s-g(s))\nonumber\\
&&Q_{11}(s)=\frac{1}{8}[ 4 q_1 \left(s^2+2\right)-\a_1^2
    \left(k_1-k_2\right){}^2 \left(s^2-3\right) g(s)^2\nonumber\\
&&-2 \a_1^2 \left(k_1-k_2\right){}^2 \log \left(s^2+1\right)+s \left(5
    \a_1^2 \left(k_1-k_2\right){}^2 s+4 \a_2\right)\nonumber\\
&&+s \left(2 \a_1 \left(-5 k_1^2+8 k_2 k_1+k_2^2\right)-4 s
    \left(k_1^2-k_2 k_1-k_2^2+q_2\right)\right)\nonumber\\
&&-2 g(s) \left(\a_1 k_1^2 \left(3 \a_1 s+s^2-5\right)+2 \a_2
    \left(s^2+1\right)\right)\nonumber\\
&&-2 \a_1 g(s) \left(k_2^2 \left(3 s \left(\a_1+s\right)+1\right)+2 k_1
    k_2 \left(4-3 \a_1 s\right)\right) ],\nonumber\\
&&Q_{22}(s)=\frac{1}{8}[ 4 q_2 \left(s^2+2\right)-\a_1^2
    \left(k_1-k_2\right){}^2 \left(s^2-3\right) g(s)^2\nonumber\\
&&+s \left(5 \a_1^2 \left(k_1-k_2\right){}^2 s-4 \a_2\right)-2 \a_1^2
    \left(k_1-k_2\right){}^2 \log \left(s^2+1\right)\nonumber\\
&&+s \left(4 s \left(k_1^2+k_2 k_1-k_2^2-q_1\right)-2 \a_1
    \left(k_1^2-4 k_2 k_1+7 k_2^2\right)\right)\nonumber\\
&&+2 g(s) \left(2 \a_2 \left(s^2+1\right)-\a_1 k_1^2 \left(3 \a_1
    s+s^2-1\right)\right)\nonumber\\
&&+2 \a_1 g(s) \left(k_2^2 \left(-3 \a_1 s+s^2+7\right)+2 k_1 k_2
    \left(3 \a_1 s+2 s^2-2\right)\right)],\nonumber\\
&&X_2(s)=\frac{1}{4}[-\a_1^2 \left(k_1-k_2\right){}^2 s^2 \log
    \left(s^2+1\right)-2 \a_1 \left(k_1-k_2\right){}^2 s\nonumber\\
&&+\a_1 \left(k_1-k_2\right){}^2 g(s) \left(\a_1 \left(s^2+1\right)
    g(s)+2 s \left(s-\a_1\right)+2\right)\nonumber\\
&&+s \left(\a_1^2 \left(k_1-k_2\right){}^2 s-2 s \left(k_1^2+k_2
    k_1+k_2^2-q_1-q_2\right)\right)],
\end{eqnarray}
where $s=z/x$ and $g(s)=\frac{\pi }{2}-2\tan^{-1}\left(1/(s+\sqrt{s^2+1})\right)$.
To rewrite the above solutions in functions of $x$ and $z$, we should consider suitable analytic continuation in order to get smooth functions at $x=0$.
In this way, we get smooth $g(z,x)$ as
\begin{eqnarray}\label{gxz}
g(z,x)=  \frac{\pi }{2}-2
\tan^{-1}\left(x/(z+\sqrt{z^2+x^2})\right).
\end{eqnarray}
Imposing NBC (\ref{NBCg}) or DBC (\ref{DBCAdS}), we can solve the integration constants
\begin{eqnarray}\label{integrationconstants3dBCFT}
\a_1=-\bar{\a}_3, \ \ \a_2=-\frac{\a_1}{2}k^2,
\end{eqnarray}
where $\bar{\a}_3$ are given by (\ref{integralconstantg1}) and (\ref{integralconstantg2}) for NBC and DBC, respectively. The BCs can not only fix the bulk solutions but also the location of $Q$ (\ref{Qhigherorder}). We obtain for both NBC and DBC
\begin{eqnarray}\label{c3Qapp1}
&&c_3=- \frac{\sinh \rho }{24}\Big[7 k_1^2+4 k_2 k_1+7 k_2^2-4
    \left(q_1+q_2\right) \nonumber\\
&&+\left(5 k_1^2+2 k_2 k_1+5 k_2^2-2 \left(q_1+q_2\right)\right) \cosh
    (2 \rho )\nonumber\\
    &&+\a_1^2 \left(k_1-k_2\right){}^2 \left((2 +\cosh (2 \rho ) )\log(
    \coth ^2 \rho )-1\right) \Big] \nonumber\\
&&+\frac{(k_1-k_2)^2}{24}  \left(\alpha _1 f_4(\rho )-1\right) \big[\alpha _1 (\cosh (2 \rho )+3)+\cosh ^3(\rho ) \coth (\rho ) \left(\alpha _1 f_4(\rho )-1\right)\big],
\end{eqnarray}
where $f_4(\rho )=\frac{\pi}{2}+2 \tan ^{-1}\left(\tanh \left(\frac{\rho }{2} \right)\right) $. It is interesting that NBC and DBC yield almost the same solutions except a different parameter $\a_1$.

Now we are ready to derive the holographic Weyl anomaly for 3d BCFT. The following approach applies to both NBC and DBC. 
 By using Einstein equations, we
can rewrite the on-shell gravitational action (\ref{action}) without gauge fields as
\begin{eqnarray}\label{action5dapp}
  I=-6\int_N \sqrt{G}
  +2\int_Q \sqrt{\gamma} (K-2 \tanh \rho).
\end{eqnarray}
To get the holographic Weyl anomaly, we need to do the integration
along $x$ and $z$, and then select the UV logarithmic divergent terms.
We divide the integration region into two parts: region I is defined
by $( z \ge 0, x \ge 0)$ and region II is defined by the complement of
region I. Let us first study the integration in region I, where only
the bulk action in (\ref{action5dapp}) contributes. Integrating along
$z$, expanding the result in small $x$ and selecting the $1/x$ term, we obtain
\begin{eqnarray}\label{actionregionI}
  I_1&=&\int_{\epsilon} dx [\frac{\pi \alpha_1^2}{2x} \text{Tr}\bar{k}^2+...]\nonumber\\
&=&\log(\frac{1}{\epsilon}) \frac{\pi \alpha_1^2}{2} \text{Tr}\bar{k}^2 + \cdots .
\end{eqnarray}
Next let us consider the integration in region II. In this case, both the bulk
action and boundary action in (\ref{action5dapp}) contribute. For the
bulk action, we first do the integral along $x$, which yields a
boundary term on $Q$. Note that since only the UV logarithmic
divergent terms are related to Weyl anomaly, we keep only the lower
limit of the integral of $x$. Adding the boundary term from bulk
integral to the boundary action in (\ref{action5dapp}), we obtain
\begin{eqnarray}\label{actionregionII}
  I_2&=& \int_{\epsilon} dz [\frac{-2\a_1-\pi \alpha_1^2-2\sinh\rho(1+\frac{\a_1}{\a_{N3}})(1+\frac{\a_1}{\a_{D3}})}{2z}\text{Tr}\bar{k}^2+...]\nonumber\\
&=&\log(\frac{1}{\epsilon}) \frac{-2\a_1-\pi \alpha_1^2}{2z}\text{Tr}\bar{k}^2 + \cdots ,
\end{eqnarray}
where we have used $\a_1=-\a_{N3}$ (\ref{integralconstantg1}) for NBC and $\a_1=-\a_{D3}$ (\ref{integralconstantg2}) for DBC above. 
Combining together (\ref{actionregionI}) and (\ref{actionregionII}) and noting that $\a_1=-\bar{\a}_3$,
we finally obtain the Weyl anomaly (\ref{3dBWA})
\begin{eqnarray}\label{HoloWeylanomaly}
\mathcal{A}=\int_{P} \sqrt{h} \; \bar{\a}_3 \text{Tr}\bar{k}^2 ,
\end{eqnarray}
with the boundary central charge $b_1=\bar{\a}_3$. Hence we obtain the universal relation (\ref{relationg}) for 3d BCFT. It should be mentioned that since we focus on solutions without $y_a$ dependence, the first term $\mathcal{R}$ of Weyl anomaly (\ref{3dBWA}) vanishes. Actually, it is easy to recover this term.  Applying the conformal map \cite{Takayanagi:2011zk}, we can obtain the holographic dual of BCFT on a round disk from the one on half space (\ref{AdSmetric},\ref{Q}). Following the above approach, we can derive the $\mathcal{R}$ term in Weyl anomaly and read off the central charge $b_1=\sinh \rho$. We remark that holographic BCFT with NBC and DBC have the same boundary central charges related to Euler densities but different boundary central charges related to extrinsic curvatures. 
 Since g-theorem in higher dimensions apply to the central charges related to Euler densities on the boundary. The holographic BCFT with NBC and DBC both obey the g-theorem. Following the same approach, we can derive the holographic Weyl anomaly for 4d BCFT and verify the  universal relation (\ref{relationg}). Since the calculations are similar to the 3d case but quite complicated, we do not repeat it here.

\section{Casimir Coefficients and Central Charges}

In the above sections, we have obtained the Casimir coefficients and central charges for holographic BCFT with NBC and DBC, respectively. In this section we study the interesting characteristics of these charges.
For simplicity we only list the main results and discuss them briefly. 

\begin{itemize}
\item[(1)] 
{\it The Casimir coefficients for NBC are greater than or equal to those for DBC.}

By``Casimir coefficients", we mean the coefficients $\alpha_d$ defined in the renormalized stress tensor (\ref{stress0}) near the boundary.
We use $\alpha_{Nd}$ and $\alpha_{Dd}$ to denote Casimir coefficients with respect to NBC and DBC, respectively. From (\ref{aN}) and (\ref{aD}), it is easy to check that 
\begin{eqnarray}\label{chargelaw1}
\bar{\a}_{Nd} \ge \bar{\a}_{Dd},
\end{eqnarray}
where the equality is saturated when the bulk boundary $Q$ is pulled back into the AdS boundary, i.e., $\rho\to \infty.$ Interestingly, $\bar{\a}_{Nd}$ gains its lower bound while $\bar{\a}_{Dd}$ gets its upper bound at $\rho\to \infty$.

\item[(2)] 
{\it The Casimir coefficients are non-negative.}

According to \cite{Miao:2018dvm}, we have $\bar{\a}_d = \lambda_d c_{nn}$, where $\lambda_d$ is a positive factor and $c_{nn}$ is the central charge defined by the two point funciton of the displacement operator
\begin{eqnarray}\label{displacementoperators}
< D^n (x) D^n(0)>=\frac{c_{nn}}{x^{2d}}.
\end{eqnarray}
From (\ref{displacementoperators}), we have $c_{nn}\ge 0$ and thus
\begin{eqnarray}\label{chargelaw2}
\bar{\a}_{Nd} \ge \bar{\a}_{Dd}\ge 0.
\end{eqnarray}
It should be mentioned that \cite{Herzog:2017kkj} find that $b_2=\frac{\pi^2}{8}c_{nn}$ for 3d BCFT and $b_4=\frac{2\pi^2}{15}c_{nn}$ for 4d BCFT.  Using the universal relation (\ref{relationg}), we get $\bar{\a}_d \ge 0$ for $d=3,4$. This can be regarded as an independent test of our claim above.

Interestingly, the positive bound is automatically satisfied by NBC. As for DBC, the bound yields $T=(d-1) \tanh\rho \ge 0$, which is reasonable if we take $T$ as the tension of the brane $Q$ \cite{Takayanagi:2011zk}.

\item[(3)] 
{\it  The Casimir coefficients for NBC reduce to the ones of free BCFT, while the Casimir coefficients for DBC vanish, when bulk boundary $Q$ is perpendicular to the AdS boundary $M$, i.e., $\rho=0$.}

From (\ref{aN}) and (\ref{aD}), we have for holographic Casimir coefficients
\begin{eqnarray}\label{chargelaw3a}
&&\lim_{\rho\to 0}\bar{\a}_{Nd}= \frac{2^{-d} \pi ^{d/2}}{(d+1) \Gamma \left(\frac{d}{2}+1\right)} C_{HT}, \\
&&\lim_{\rho\to 0}\bar{\a}_{Dd}= 0, \label{chargelaw3b}
\end{eqnarray}
where $C_{HT}$ is the holographic central charge \cite{Buchel:2009sk}
\begin{eqnarray}\label{CT}
C_{HT}=\frac{2(d+1)}{d-1}\frac{\Gamma[d+1]}{\pi^{d/2}\Gamma[d/2]}.
\end{eqnarray}
In general the central charge $C_T$ is defined by the two point functions of stress tensor far away from the boundary
\begin{eqnarray}\label{twopoint}
<T_{\mu\nu}(x)T_{\lambda\rho}(0)>=\frac{C_T}{|x|^{2d}}I_{\mu\nu,\lambda\rho}(x)
\end{eqnarray}
with $I_{\mu\nu,\lambda\rho}$ a dimensionless tensor fixed by symmetry. 

It is remarkable that free BCFTs seem to obey the same relation as (\ref{chargelaw3a}),
\begin{eqnarray}\label{chargelawfree}
\bar{\a}_{Fd}= \frac{2^{-d} \pi ^{d/2}}{(d+1) \Gamma \left(\frac{d}{2}+1\right)} C_{FT}
\end{eqnarray}
where $\bar{\a}_{Fd}$ and $C_{FT}$ are the Casimir coefficients defined by (\ref{stress0})  and the central charge defined by (\ref{twopoint}) for free BCFTs,  respectively.  One can verify  (\ref{chargelawfree})  by free BCFTs in three and four dimensions \cite{Fursaev:2015wpa,Deutsch:1978sc,Jensen:2015swa} and by free scalars in general dimensions \cite{Miao:2018dvm}.  It is expected that (\ref{chargelawfree}) applies to general free BCFTs in general dimensions.  

To end this section, let us draw the figures of Casimir coefficients for NBC and DBC in three and four dimensions, respectively. See figure 2 and figure 3. 
From these figures, we learn that Casimir coefficients for NBC are indeed greater than or equal to those for DBC. And the Casimir coefficients are indeed non-negative for physical tension $T=(d-1) \tanh\rho \ge 0$. 
\begin{figure}[t]
\centering
\includegraphics[width=7cm]{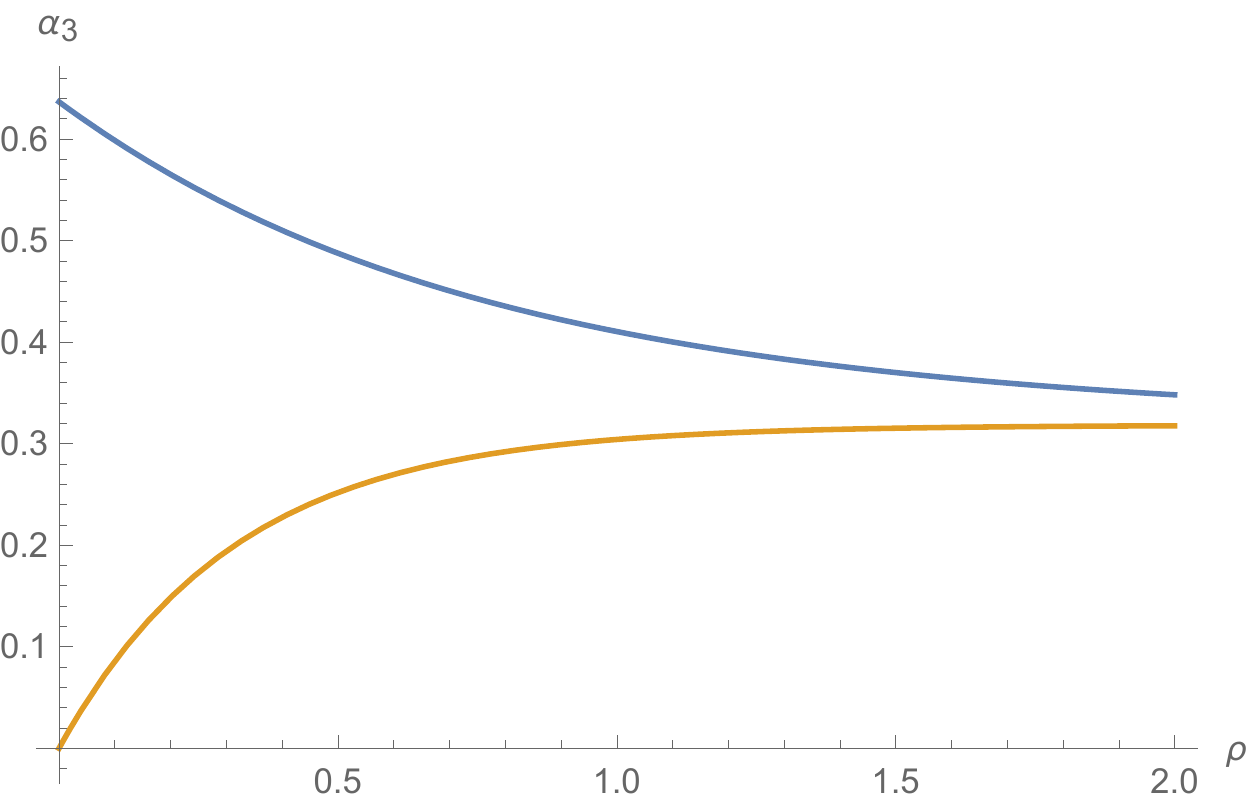}
\caption{3d Casimir coefficients for NBC (blue line) and DBC (yellow line).}
\end{figure}
\begin{figure}[t]
\centering
\includegraphics[width=7cm]{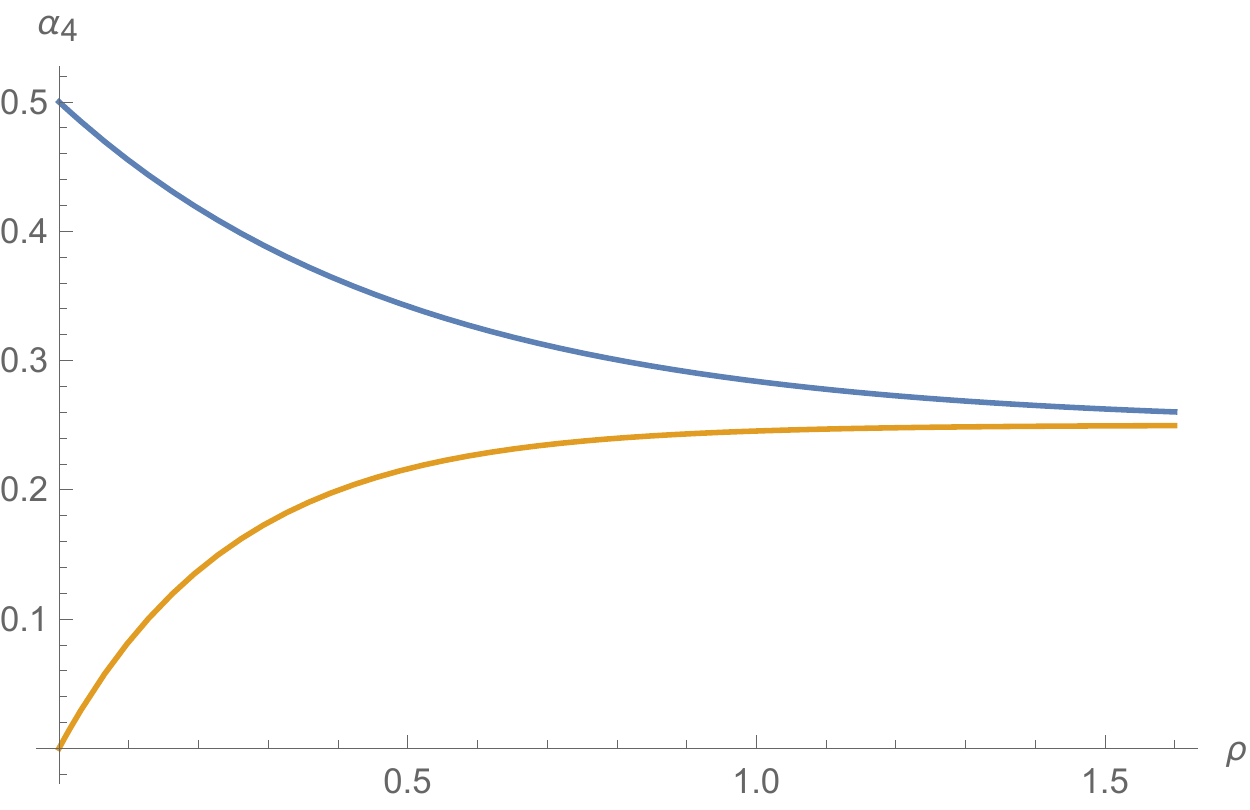}
\caption{4d Casimir coefficients for NBC (blue line) and DBC (yellow line).}
\end{figure}
\end{itemize}

\section{Conclusions and Discussions}

In this paper, we have investigated the holographic BCFT with Dirichlet BC and find it works as well as the one with Neumann BC. For example, AdS is a vacuum solution to Dirichlet BC and the g-theorem is obeyed by this theory. Furthermore, it gives the expected one point function of the stress tensor (current) and the universal relations between the stress tensor (current) and Weyl anomaly. It is remarkable that the boundary central charge related to B-type Weyl anomaly, or equivalently, the Casimir coefficient for Dirichlet BC is less than or equal to the one for Neumann BC.  We have also studied the relative BC for gauge fields, which is the counterpart of Dirichlet BC for gravitational fields. We find an exact solution to this BC, which implies that a constant magnetic field in the bulk can induce a constant current on the boundary in three dimensions. And the boundary current gets the maximum value at zero temperature. It is interesting to measure this boundary effect in laboratory. In this paper, we discuss only the tip of the iceberg for Dirichlet BC, i.e., we fix the boundary metric to be that of AdS. It is natural to study more general metrics on $Q$, such as the ones describing gravitational waves and black holes. Besides, one can generalize the discussions of this paper to other fields, such as scalars and higher spin fields. 
Finally, it is also interesting to study other kinds of BCs for holographic BCFT, such as the conformal BC or mixed BC. We notice that the conformal BC is more subtle, which is less restrictive than Dirichlet BC and Neumann BC. We leave a careful study of this problem to future work.

\section*{Acknowledgements}
We would like to thank Chong-Sun Chu, Shu Lin, Jia-Rui Sun and Mitsutoshi Fujita
for useful discussions and comments. This work is supported by the funding of Sun Yat-Sen University.

\appendix

\end{document}